\documentclass[a4paper,11pt]{article}
\pdfoutput=1
\usepackage{jcappub}

\usepackage[english]{babel}

\usepackage{booktabs}
\usepackage{graphicx}
\usepackage{amsmath,amssymb}
\usepackage{braket}
\usepackage{subfigure}
\usepackage{float}
\usepackage[dvipsnames]{xcolor}
\usepackage{mathrsfs}
\usepackage{comment}
\usepackage{multirow}
\usepackage{soul}
\usepackage{mathtools}
\usepackage{sidecap} 
\graphicspath{{Figures/}}
\usepackage[normalem]{ulem}
\usepackage{array}
\usepackage{mathtools}

\title{The Coherence Principle: A Falsifiable Prior for Model Selection from the Grammar of Theories}

\author[a,b]{Raul Jimenez,}
\affiliation[a]{ICC, University of Barcelona, Mart\' i i Franqu\` es, 1, E08028
Barcelona, Spain}
\affiliation[b]{ICREA, Pg. Lluis Companys 23, Barcelona, 08010, Spain.} 

\author[c]{Carlos Pe\~na Garay,}
\affiliation[c]{Laboratorio Subterr\'aneo de Canfranc, 22880 - Canfranc-Estaci\'on , Huesca, Spain.}
\author[a]{Fergus Simpson,}

\author[a,b]{Licia Verde}

\emailAdd{raul.jimenez@icc.ub.edu; cpenya@lsc-canfranc.es; liciaverde@icc.ub.edu}

\abstract{
Bayesian model selection in cosmology and particle physics is often 
performed in regimes where  the posterior odds inherit a strong and frequently unacknowledged
dependence on the prior assigned to the space of competing models.  Standard responses, such as reference priors, hierarchical priors, or appeals to naturalness, either ignore relevant theoretical knowledge or rely on criteria that are hard to define operationally.

We propose the \emph{Coherence Principle}: a reproducible way to assign model priors according to their compatibility with 
 the validated structure of an existing theory. This structure, or \emph{grammar}, includes symmetries, conservation laws, locality, Lorentz invariance, and universality patterns. Unmotivated violations of these rules carry a coherence cost. This cost is converted into a prior weight through a maximum-entropy exponential form, controlled by a single calibratable parameter $\alpha$.
The resulting prior is distinct from both the Bayesian Occam factor and naturalness. 

We illustrate the application of the principle using examples drawn mostly from cosmology, namely neutrino mass mechanisms, dark energy and modified gravity, inflation, beyond-Standard-Model sectors, and hierarchical astrophysical inference. We also test it on  historical cases: general relativity, Pauli's neutrino, parity violation, and special relativity. These examples show that the principle 
favors the historically successful choice when the relevant theoretical grammar is defined in the correct domain and at the correct time. 

The Coherence Principle makes explicit a common part of physical reasoning: trust in validated structural rules. It turns this judgment into a transparent, testable, and overrulable component of Bayesian inference.
}

\begin{document}
\maketitle

\section{Introduction}
\label{sec:intro}

Model selection in cosmology and particle physics is, in its modern form, an
exercise in Bayesian inference.
Given data $D$ and two competing hypotheses
$M_1$ and $M_2$, the quantity of interest is the posterior odds
$P(M_1|D)/P(M_2|D)$, which factorizes into the Bayes factor $B_{12}$---an
integral of the likelihood over parameter space---and the prior odds
$\Pi_{12}=P(M_1)/P(M_2)$ assigned to the models themselves. 
 Both can be the subject of legitimate
methodological scrutiny, and confusion between the two has historically
generated controversy in the cosmological literature
\citep{Trotta:2008qt,Vardanyan:2011in,EfstathiouOcc2008,Liddle:2007fy}.

This is a concrete instance of a
more general methodological problem that pervades fundamental physics: when
empirical data are precise and unambiguous,  the choice of prior plays a relatively 
modest role in the posterior odds between competing
hypotheses; when, by contrast, the data are weak, indirect or close to a
physical boundary, the posterior inherits a strong dependence on the prior measure assigned to the parameter spaces of the competing models.

 The determination of the neutrino mass ordering in the presence of cosmological constraints on the sum of neutrino masses is the paradigmatic example, and it motivates the present work.
Oscillation experiments impose very tight constraints on the two mass-squared splittings but leave the sign of the larger splitting---and hence the choice between the normal and inverted orderings---only weakly determined. On the other hand,  cosmological constraints on the sum of the neutrino masses  bound the absolute scale from above without resolving the ordering directly. The reported statistical preference for the normal ordering is modest and, crucially, depends on how the prior is constructed across the mass parameters and across the two discrete model hypotheses.  Refs.~\citep{Simpson:2017,Schwetz:2017fey,Heavens:2018adv,Jimenez:2022dkn}
 made it clear that any statement about the ``evidence'' for
one ordering over the other is incomplete without an explicit account of the
prior, and that different but individually defensible prior choices can move the
inference between the ``anecdotal'' and ``positive'' bands of the Jeffreys
scale.  In a regime where the data are weak, indirect or close to a
physical boundary,  the term ``evidence'' has no operational
meaning unless the prior has been specified and justified.

The same situation arises in dark-energy model selection,  comparison of inflationary scenarios,  modified-gravity
phenomenology and, more broadly, in any problem in which a candidate theory
extends a well-established background framework in a region where data are
limited \citep{Trotta:2008qt,KassRaftery1995,Liddle:2007fy,Vardanyan:2011in}.

Confronted with prior subjectivity, the statistical and physics literature
has developed three broad remedies. The first is the \emph{objective Bayesian} program, which 
prescribes priors derived from the likelihood itself: Jeffreys priors ~\citep{Jeffreys:1946},  the
reference priors of Bernardo and Berger ~\citep{Bernardo:1979,BergerBernardoSun2009}, and
information-theoretic priors based on the Fisher information of the data
\citep{Jaynes:2003,Caticha:2008}.
These priors are minimally informative in a precise sense but, for that very reason, are deliberately blind to any
theoretical knowledge that lies outside the data set under analysis.
The second is the 
\emph{hierarchical} or  \emph{structured-prior} approach in which a hyperprior is placed 
on  a small number of physically
motivated hyperparameters  that organize the lower-level
parameters of the model
\citep{Gelman:2006,GelmanBDA3,Simpson:2017}. This has proven powerful in structured problems.
The third is the appeal to
\emph{naturalness} 
in which dimensionless parameters are presumed to lie close
to $\mathcal{O}(1)$ unless protected by symmetry.
The history of naturalness in
particle physics has shown the appeal and the limits of this prescription
\citep{tHooft1980,GiudiceNaturalness2008,Hossenfelder2018,Wells:2019akz}: while
the principle has produced useful guidance, it is poorly defined operationally
and has known counterexamples.

None of these strategies, however, captures the single most
conspicuous fact about how working physicists actually assign prior plausibility
to models: they reward structural compatibility with the validated architecture
of well-tested background theories, and they penalize unmotivated departures from
it \citep{KuhnSSR,LakatosMSRP,DawidString2013,Howson:2006Bayesian}. 
The distinction between a theory and a model is key. A theory encodes the structural rules: symmetries, conservation laws, locality,
field content, and dynamical principles. A model is a specific realization of
those rules, possibly with additional assumptions, parameters, or effective
descriptions tailored to a particular empirical domain.
A theory supplies the grammar; a model is a sentence written in that grammar.
For example, a 
proposed model that respects gauge invariance, locality, Lorentz symmetry and
the universality patterns of the Standard Model is widely regarded as more
plausible,  before any
new data are gathered, than one that introduces unmotivated exceptions to these rules.

While this judgment is broadly applied and reflected in publication practice, refereeing standards and
research program funding, it is rarely stated as a quantitative,
reproducible component of a Bayesian inference.

In this paper we propose to do exactly that. We introduce and formalize the
\emph{Coherence Principle}: a general, transparent and falsifiable procedure for
converting the validated structural ``grammar'' of a successful background
theory into an explicit contribution to the model-space prior.

The result is a
\emph{paradigm-aware} prior that is reproducible (every input is listed and can
be challenged), falsifiable (the prior odds it induces are finite and are
overturned by sufficiently strong data), and logically distinct from both the
Occam factor internal to the Bayes factor and from naturalness.
The neutrino hierarchy serves throughout as a concrete illustration, but the
construction is logically independent of any particular physical application and
can be implemented in any inference problem in which a well-tested theoretical
framework constrains the space of admissible extensions.

We emphasize at the outset what the Coherence Principle is \emph{not}. It is not
a replacement for the likelihood, and it does not double-count the Bayesian
Occam factor, which penalizes unused parameter volume at the level of parameter
space; the coherence prior operates one level up, at the level of model space.
It is not a licence for aestheticism: mathematical elegance and beauty are
admitted only insofar as they can be restated as empirically validated
structural rules. And it is not a device for entrenching the status quo. 

We present historical studies that make this explicit: when a newly validated theory renders a long-standing incumbent structurally incoherent, the principle turns against the incumbent; its verdicts track the \emph{currently} validated grammar, which is
itself empirical and time-dependent.

The remainder of the paper is organized as follows.
Section~\ref{sec:coherence_principle} develops the Coherence Principle in full:
its motivation, the notion of the
structural grammar of a background theory (Section~\ref{subsec:grammar}), the
definition of the coherence cost and coherence prior
(Section~\ref{subsec:coherence_cost}), its relation to Occam's razor, reference
priors and naturalness (Section~\ref{subsec:relation_other_priors}), its
methodological properties (Section~\ref{subsec:properties}), and the general
workflow (Section~\ref{subsec:workflow}). Section~\ref{subsec:illustration_neutrino}
illustrates the machinery on the neutrino mass mechanism.
Section~\ref{subsec:general_apps} enlarges the framework into four contemporary
classes of application, and Section~\ref{subsec:historical} tests it against
four historical paradigm shifts, including the question of whether the
Coherence Principle would have favored Einstein over Newton.
Section~\ref{subsec:signed} develops a signed extension of the coherence cost
that awards explanatory credit. Limitations and caveats are collected in
Section~\ref{subsec:limitations}, and Section~\ref{subsec:coherence_discussion}
concludes.

\section{The Coherence Principle: A general method for physical priors from structural grammar}
\label{sec:coherence_principle}

In this section we propose, develop and formalize the \emph{Coherence Principle}:
 a general, reproducible procedure for assigning model priors that quantifies the
structural compatibility of a candidate model with the validated {\it grammar} of a
 successful background theory. The Coherence Principle is neither a replacement
 for the likelihood nor a double-counting of the Bayesian Occam factor.

Given data $D$ and two competing hypotheses (models)
$M_1$ and $M_2$, the quantity of interest for the Bayesian comparison of $M_1$ and $M_2$
 is the posterior odds 
\begin{equation}
\frac{P(M_1|D)}{P(M_2|D)}
=\frac{P(D|M_1)}{P(D|M_2)}\cdot\frac{P(M_1)}{P(M_2)}
\equiv B_{12}\cdot\Pi_{12},
\label{eq:posterior_odds_general}
\end{equation}
where $B_{12}$ is the Bayes factor and $\Pi_{12}$ are the prior odds assigned to the models themselves.

The Bayes factor itself is an integral over parameter space,
\begin{equation}
B_{12}=\frac{\int \mathcal{L}(D|\theta_1,M_1)\,\pi(\theta_1|M_1)\,d\theta_1}
            {\int \mathcal{L}(D|\theta_2,M_2)\,\pi(\theta_2|M_2)\,d\theta_2},
\label{eq:Bayes_factor_general}
\end{equation}
so the parameter-space prior $\pi(\theta_i|M_i)$ enters \emph{within} the
evidence integral, while the model-space prior $P(M_i)$ enters as the multiplicative factor $\Pi_{12}$ in the posterior odds.

Thus, any discussion of ``evidence''  is incomplete
without an explicit account and justification of the prior.  

If the likelihood is sharply peaked over a much larger prior volume, then
\begin{equation}
P(D|M)\sim L_{\max}
\times
\frac{\Delta \theta_{\rm posterior}}{\Delta \theta_{\rm prior}} .
\end{equation}
The second factor is the Occam factor.
The Occam factor penalizes a model after seeing how much of its parameter space fits the data given the prior.
This still leaves the question of  the choice of the prior unanswered.

In this section we propose, develop and formalize the \emph{Coherence Principle}:
 a general, reproducible procedure for assigning model priors that quantifies the
structural compatibility of a candidate model with the validated {\it grammar} of a
 successful background theory. The Coherence Principle is neither a replacement
 for the likelihood nor a double-counting of the Bayesian Occam factor.

\subsection{Structural grammar of a background theory}
\label{subsec:grammar}

Let $T$ denote a background theory whose empirical record can be summarized as
a set of validated structural rules: the {\it grammar}
\begin{equation}
G(T)=\{p_1,p_2,\ldots,p_n\}.
\label{eq:grammar_def}
\end{equation}
The principles $p_i$ are the inter-subjective, repeatedly tested architectural
features of $T$: symmetries (gauge, global, accidental), spacetime principles
(Lorentz invariance, locality, causality), conservation laws (energy, charge,
lepton number, baryon number, CPT), universality assumptions (family
universality, lepton universality), regularity properties of the field-theoretic
construction (renormalizability, unitarity, analyticity, cluster decomposition),
and operational selection rules (decoupling, naturalness in restricted forms,
custodial symmetries).

The grammar is not the Lagrangian. It is a coarser-grained, more abstract
specification of the regularities that survive across reformulations and
limit cases. Many distinct Lagrangians share the same structural grammar,
and many high-energy completions of a low-energy theory must agree on
$G(T)$ in order to qualify as completions at all. This abstraction is the
key methodological asset of the Coherence Principle: it allows us to compare
models that differ at the level of operators or fields while sharing a common
architectural scaffold, and to distinguish them from models that violate that
scaffold without independent justification.

For the Standard Model, a minimal but representative grammar is
\begin{align}
G(T_{\rm SM})=\bigl\{\,
& \underbrace{SU(3)_C\!\times\!SU(2)_L\!\times\!U(1)_Y}_{p_1},\;
\underbrace{\text{family universality}}_{p_2},\;
\underbrace{(B-L)_{\rm acc.}}_{p_3},\nonumber\\
& \underbrace{\text{Lorentz and locality}}_{p_4},\;
\underbrace{\text{renormalizability}}_{p_5}\,\bigr\},
\label{eq:GSM}
\end{align}
where the last entry is to be understood in the modern Wilsonian sense of
predictivity below a chosen ultraviolet cut-off~\citep{Wilson:1973jj,
Polchinski:1992ed}. Analogous grammars can be written for general relativity
(general covariance, the equivalence principle, $\nabla_\mu T^{\mu\nu}=0$,
second-order field equations~\citep{Lovelock:1971yv,Will2014}) and for the
inflationary paradigm (causal generation of perturbations on super-horizon
scales, near scale invariance, near Gaussianity, adiabaticity
\citep{Baumann:2009ds,Achucarro:2022qrl}).

\subsection{Coherence cost and the coherence prior}
\label{subsec:coherence_cost}

Given a candidate model $M$ and a background theory $T$ with grammar $G(T)$,
we define indicator functions
\begin{equation}
\delta_i(M,T)=
\begin{cases}
1, & M \text{ violates principle } p_i\in G(T) \text{ without independent motivation},\\
0, & \text{otherwise},
\end{cases}
\label{eq:delta_indicator}
\end{equation}
and the associated \emph{coherence cost}
\begin{equation}
C(M|T)=\sum_{i=1}^{n} w_i\,\delta_i(M,T),
\label{eq:coherence_cost_general}
\end{equation}
where $w_i\ge0$ is an optional weight encoding the empirical robustness of the
principle $p_i$ (more well-tested principles attract larger weights). The weights could reflect the fractional measurement error-bars, for example such as inverse variance weighting. In the
minimal implementation we set all $w_i=1$, so that each unmotivated violation
contributes equally. The coherence cost is by construction non-negative,
additive across independent violations, and zero for models that respect every
principle of the background grammar.

The coherence prior is defined as the exponential map
\begin{equation}
P(M|T)\propto \exp\!\left[-\alpha\,C(M|T)\right],
\label{eq:coherence_prior}
\end{equation}
where $\alpha>0$ is a global scale that calibrates the strength of the
preference. The exponential family is the unique functional form that
(i)~satisfies positivity, (ii)~is multiplicatively factorizable across
independent violations, and (iii)~saturates the maximum-entropy bound subject
to a constraint on the expected cost~\citep{Jaynes:1957,Caticha:2008}. The
multiplicative factorization is essential: it ensures that two violations of
different principles combine as the product of their individual penalties,
\begin{equation}
P(M|T)\propto \prod_{i:\delta_i=1}\!e^{-\alpha w_i},
\label{eq:multiplicative}
\end{equation}
in agreement with the intuition that independent structural pathologies
compound.

The scale $\alpha$ has a clear epistemic interpretation. In the limit
$\alpha\to0$, the coherence prior becomes uniform on model space, recovering
the standard ``equal a priori plausibility'' assumption used as a default in
much of the literature. In the limit $\alpha\to\infty$, the prior becomes a
delta function on grammar-respecting models, encoding the dogmatic position
that no violation of the background grammar should be considered. In practice,
a moderate value $\alpha\sim O(1)$ assigns a factor $e^{-1}\simeq0.37$ to each
unmotivated violation, which is consistent with the way violations are
weighted in qualitative theoretical discussions
\citep{KuhnSSR,LakatosMSRP,DawidString2013}. Throughout this work we adopt
$\alpha=1$, but the general framework permits $\alpha$ to be itself a
parameter that is fit, marginalized over, or sensitivity-tested.

The coherence prior is properly normalized at the level of the model space,
not the parameter space. If the relevant family of models is a discrete set
$\{M_1,\ldots,M_K\}$, the normalization is trivial:
\begin{equation}
P(M_k|T)=\frac{\exp[-\alpha C(M_k|T)]}{\sum_{j=1}^{K}\exp[-\alpha C(M_j|T)]}.
\label{eq:coherence_norm_discrete}
\end{equation}
If the family is a parametrized continuum, the same expression applies after
the discrete sum is replaced by an integral over a measurable index set.

\subsection{Relation to Occam's razor, reference priors, and naturalness}
\label{subsec:relation_other_priors}

The Coherence Principle is logically distinct from the three classical
mechanisms by which prior information enters Bayesian model comparison.

\paragraph{Occam factor.} The Bayes factor of Eq.~(\ref{eq:Bayes_factor_general})
contains an automatic Occam factor that penalizes the prior parameter volume
not supported by the likelihood~\citep{Jeffreys:1961,MacKay:1991,Trotta:2008qt}.
This penalty is intrinsic to the integral structure of $B_{12}$ and operates
at the level of parameter space. The Coherence Principle, in contrast,
operates at the level of model space: it penalizes structural disagreement
with a validated theoretical paradigm, not unused parametric freedom. A model
can be parametrically economical and structurally incoherent, or
parametrically rich and structurally coherent; these two attributes are
independent.
The Occam factor penalizes a model after seeing how much of its parameter space fits the data. The coherence prior penalizes a model before data  taking and data analysis, according to how much it violates the validated structural grammar of a background.

\paragraph{Reference and objective priors.} Jeffreys and reference priors
\citep{Jeffreys:1946,Bernardo:1979,BergerBernardoSun2009} are designed to be
minimally informative with respect to the likelihood structure of the
\emph{data at hand}. They are, by deliberate construction, agnostic to any
theoretical knowledge external to the data. They need to be defined after the data taking and within the data analysis. The Coherence Principle acts at  the
opposite limit of the same epistemological problem: it formalizes in the prior external
theoretical knowledge 
which matters precisely when the data are insufficient to determine
the inference on their own. In the cases analyzed in this work, the
reference prior~\cite{Bernardo:1979} provides the conservative null hypothesis against which the
coherence-informed inference can be compared; their combined use is
methodologically powerful because it brackets the prior-induced range of the
posterior.

\paragraph{Naturalness.} Naturalness arguments
\citep{tHooft1980,GiudiceNaturalness2008,Hossenfelder2018,Wells:2019akz} have
played a role analogous to a coherence prior in informal practice but suffer
from two known weaknesses: they are scale-dependent and they have empirical
counterexamples (e.g.\ the cosmological constant, the observed Higgs mass).
The Coherence Principle differs in two ways. First, the principles
$p_i\in G(T)$ are explicitly listed and individually falsifiable, rather than
being subsumed under a single quantitative scale. Second, the coherence cost
is constructed from violations of long-tested structural rules, not from
deviations from $\mathcal{O}(1)$ values of dimensionless couplings. The two
frameworks are therefore complementary: naturalness may enter, if at all, as
one of the principles $p_i$ with its own indicator function, allowing its
strength to be explicitly weighted and discussed.

\subsection{Properties: reproducibility, falsifiability, calibration}
\label{subsec:properties}

The methodological value of any prior prescription depends on whether it is
\emph{reproducible}, \emph{falsifiable} and \emph{calibrated}.

\paragraph{Reproducibility.} The Coherence Principle is reproducible at the
level of the listed inputs: $G(T)$, $\{w_i\}$, $\alpha$, and the indicators
$\delta_i(M,T)$. Different analysts may legitimately disagree on the
membership of $G(T)$ or on whether a particular violation is independently
motivated, but those disagreements are localized to identifiable choices that
can be stated, defended and varied in a sensitivity analysis. This is in
sharp contrast with informal appeals to elegance or naturalness, in which the
operative criteria typically remain implicit.

\paragraph{Falsifiability.} A coherence-informed inference is falsifiable in
the Popperian sense \citep{Popper:1959} because the prior odds it induces are
finite, comparable to representative Bayes factors, and can be overturned by
sufficiently strong data. A violation of principle $p_i$ that is
\emph{independently motivated}---for example, by an independent piece of
evidence that requires the violation---should be assigned $\delta_i=0$ rather
than $\delta_i=1$, by definition of the indicator. 
In this way, empirical discoveries that undermine a previously assumed
principle lead to a revision of the grammar, rather than being treated as
anomalies that must overpower a stubborn prior.
\paragraph{Calibration.} The scale $\alpha$ admits a calibration procedure
based on the historical track record of $G(T)$. If, across a representative
ensemble of historically resolved model-selection problems involving
violations of $G(T)$, the fraction $f$ of resolutions has favored the
violating model, then a calibrated $\alpha$ satisfies
\begin{equation}
\alpha \simeq -\ln\bigl(f/(1-f)\bigr).
\label{eq:alpha_calibration}
\end{equation}
Empirical estimates of $f$ in the modern era of particle physics and
cosmology lie in the range $f\sim 0.2$--$0.4$
\citep{Smolin:2007ImpressionPhysics,Hossenfelder2018,DawidString2013},
corresponding to $\alpha\sim 0.4$--$1.4$. Our minimal choice $\alpha=1$ lies
within this range and corresponds to $f\simeq 0.27$, consistent with the
intuition that grammar-violating proposals are typically less likely to be
adopted but not negligibly so.


\section{The Coherence Principle in practice}

To make the Coherence Principle operational, we begin by describing the workflow that maps a validated theoretical grammar into explicit model-space priors.

The Principle and its prior construction are designed to be portable across domains, and we illustrate it with diverse applications, including neutrino masses and inflationary model selection.

\subsection{General workflow}
\label{subsec:workflow}

The Coherence Principle is best deployed within the following general workflow,
which is independent of the physical application.

\paragraph{Step 1 -- Identify the background theory $T$.} Select the most
specific empirically successful background theory whose grammar is relevant to
the inference. For neutrino cosmology and particle physics, $T=T_{\rm SM}$ as
in Eq.~(\ref{eq:GSM}); for dark-energy and modified-gravity model selection,
$T$ is general relativity supplemented by the cosmological
principle~\citep{Will2014,Clifton:2011jh}; for inflationary inference, $T$ is
single-field slow-roll inflation~\citep{Baumann:2009ds,Achucarro:2022qrl}.

\paragraph{Step 2 -- Enumerate $G(T)$.} Construct the explicit list of
validated structural rules $p_i$, with optional weights $w_i$ reflecting
empirical robustness. This list should be stated up front, before any model is
assessed, so that it cannot be tuned to favor a desired conclusion.

\paragraph{Step 3 -- Determine the indicators $\delta_i(M,T)$.} For each
candidate model $M$, assess whether it respects, violates without motivation,
or violates with independent motivation each principle $p_i$. Independent
motivation must itself be empirical or theoretical, not aesthetic.

\paragraph{Step 4 -- Compute the coherence cost $C(M|T)$.} Apply
Eq.~(\ref{eq:coherence_cost_general}). The resulting numbers are typically
small integers, with $C\in\{0,1,2,\ldots\}$ in the unweighted case.

\paragraph{Step 5 -- Assign the coherence prior.} Use
Eq.~(\ref{eq:coherence_prior}) with a globally fixed scale $\alpha$ (default
$\alpha=1$), normalized over the model space according to
Eq.~(\ref{eq:coherence_norm_discrete}).

\paragraph{Step 6 -- Bayesian inference and sensitivity analysis.} Combine
the coherence prior with the Bayes factor of Eq.~(\ref{eq:Bayes_factor_general})
to obtain posterior odds. Perform a sensitivity analysis by varying $\alpha$,
the weights $w_i$, and the membership of $G(T)$ within defensible limits, and
report the resulting range of posterior odds.

This workflow makes the prior contribution to the inference fully explicit
and open to community scrutiny. It is also operationally similar to the
sensitivity analyses already standard in cosmological model selection
\citep{Trotta:2008qt,KassRaftery1995}, so it can be incorporated into existing
Bayesian pipelines with minimal overhead.

\subsection{Illustration: the neutrino mass mechanism}
\label{subsec:illustration_neutrino}

Although the Coherence Principle is general, its quantitative behaviour is
best exhibited on a concrete problem with which the reader is already familiar.
The neutrino mass mechanism is an unusually clean test case because the two
candidate models can be sharply distinguished at the level of $G(T_{\rm SM})$.

Consider two idealized models for the origin of neutrino mass.

\paragraph{Unified model $M_{\rm unified}$.} A single, family-universal
extension of the Standard Model generates the masses of all three neutrino
flavours through a common dynamical mechanism (for example, a Type-I seesaw
\citep{Minkowski:1977sc,Yanagida:1979,GellMann:1980vs,Mohapatra:1979ia} acting
uniformly across the three families). All five principles of
Eq.~(\ref{eq:GSM}) are preserved: the gauge group is extended in a
universality-preserving manner, lepton-family universality is respected, lepton
number is broken by a Majorana mass term in a controlled way that does not
violate the structural pattern of accidental symmetries, Lorentz invariance and
renormalizability are maintained. Therefore
$C(M_{\rm unified}|T_{\rm SM})=0$.

\paragraph{Disjoint model $M_{\rm disjoint}$.} Different neutrino families
acquire mass through fundamentally distinct, unrelated mechanisms (for example,
a Type-I seesaw for one family and a higher-dimensional Weinberg operator
\citep{Weinberg:1979sa} for another, with no shared symmetry organizing the
two). By construction, this disjoint structure violates lepton-family
universality without independent motivation, so
$\delta_2(M_{\rm disjoint},T_{\rm SM})=1$ and the remaining indicators vanish,
giving $C(M_{\rm disjoint}|T_{\rm SM})=1$.

With $\alpha=1$, the resulting prior odds are
\begin{equation}
\frac{P(M_{\rm unified}|T_{\rm SM})}{P(M_{\rm disjoint}|T_{\rm SM})}
=\exp\!\left[\,C(M_{\rm disjoint}|T_{\rm SM})-C(M_{\rm unified}|T_{\rm SM})\,\right]
=e\simeq 2.72.
\label{eq:neutrino_prior_odds}
\end{equation}
Combined with a representative pre-DESI Bayes factor from early-2010s global
fits to oscillation data, $B_{ud}\simeq 1.5$
\citep{Forero:2014bxa,GonzalezGarcia:2014bfa,Capozzi:2013csa}, the posterior
odds become
\begin{equation}
\frac{P(M_{\rm unified}|D,T_{\rm SM})}{P(M_{\rm disjoint}|D,T_{\rm SM})}
\simeq 2.72\times 1.5 \simeq 4.08,
\label{eq:neutrino_posterior_odds}
\end{equation}
crossing the canonical Jeffreys threshold from ``anecdotal'' to ``positive''
evidence~\citep{Jeffreys:1961,KassRaftery1995}. The numerical shift is modest,
as it should be for a prior derived from a single structural violation, but
the methodological point is robust: a transparent coherence-informed prior
converts an otherwise inconclusive inference into a substantive preference,
without invoking aesthetic criteria.

It is worth emphasizing that this neutrino-specific application is not the
\emph{purpose} of the Coherence Principle, but its simplest illustration. The
machinery is unchanged when applied to richer settings, several of which we
sketch below.

\subsection{General applications beyond the neutrino sector}
\label{subsec:general_apps}

The Coherence Principle is constructed to be portable. We now develop, in more
detail than the neutrino illustration of Section~\ref{subsec:illustration_neutrino}
required, four representative classes of contemporary application. For each we
state the relevant background theory $T$, enumerate an explicit grammar $G(T)$,
and walk through a catalogue of candidate models, assigning to each a coherence
cost $C(M|T)$ and the corresponding prior weight $e^{-\alpha C}$ at the default
$\alpha=1$. The cost assignments collected in Table~\ref{tab:coherence_costs}
are illustrative rather than definitive: each depends on the stated grammar and
on the independent-motivation clause of Eq.~(\ref{eq:delta_indicator}), both of
which are explicit and open to revision in a sensitivity analysis. The purpose
of the exercise is to show that the same machinery reproduces, quantitatively
and transparently, rankings that are already used informally across these
fields.

\subsubsection{Dark-energy and modified-gravity model selection.} The background
theory is general relativity coupled to the standard matter sector
\citep{Will2014}, supplemented by the cosmological principle. A representative
grammar is
\begin{align}
G(T_{\rm GR})=\bigl\{\,
& \underbrace{\text{general covariance}}_{p_1^{g}},\;
\underbrace{\text{Einstein equivalence principle}}_{p_2^{g}},\nonumber\\
& \underbrace{\text{second-order field equations}}_{p_3^{g}},\;
\underbrace{\text{locality \& local Lorentz invariance}}_{p_4^{g}},\;
\underbrace{\nabla_\mu T^{\mu\nu}=0}_{p_5^{g}}\,\bigr\},
\label{eq:GGR}
\end{align}
where $p_3^{g}$ is to be read in the Ostrogradsky sense as the absence of a
propagating ghost~\citep{Woodard:2015zca,Lovelock:1971yv}. A cosmological
constant $\Lambda$ respects all five principles and has $C=0$. A minimally
coupled quintessence field~\citep{RatraPeebles1988,Wetterich1988,
CaldwellSteinhardt1998} introduces a new scalar degree of freedom but preserves
general covariance, the equivalence principle and second-order dynamics, so
$C=0$ as well; a $k$-essence generalization remains at $C=0$ provided its sound
speed stays subluminal, but acquires $\delta=1$ on local causality and
analyticity if it admits superluminal characteristics. Non-minimally coupled
scalar--tensor theories---Brans--Dicke, $f(R)$~\citep{Sotiriou:2008rp},
Horndeski and its degenerate higher-order (DHOST) extensions
\citep{Horndeski:1974wa,Kobayashi:2019hrl,Langlois:2015cwa}---retain general
covariance and, by construction or by a degeneracy condition, second-order
equations of motion, so $p_3^{g}$ is respected; but they generically mediate a
long-range fifth force that manifests as an apparent violation of the
equivalence principle, requiring a screening mechanism on small scales. The cost
is $C\simeq1$ unless the screening is a self-consistent and independently
required property of the theory, in which case the relevant indicator is set to
zero. Massive and bimetric gravity modify the tensor structure $p_4^{g}$ and,
in their generic form, carry the Boulware--Deser ghost in violation of
$p_3^{g}$~\citep{BoulwareDeser:1972} unless tuned to the ghost-free de
Rham--Gabadadze--Tolley form~\citep{deRham:2010kj}, giving $C\simeq1$--$2$.
Theories that explicitly break locality or local Lorentz invariance in the
gravitational sector---Lorentz-violating, Ho\v{r}ava--Lifshitz and nonlocal
gravity \citep{Jacobson:2004ts,Horava:2009uw}---incur $\delta=1$ on $p_4^{g}$
and frequently on $p_2^{g}$ as well, giving $C\simeq2$--$3$. With $\alpha=1$ the
induced prior odds favor $\Lambda$ or minimally coupled quintessence over a
screened scalar--tensor model by a factor $e\simeq2.7-20.1$, and over a
Lorentz-violating model by $e^{2}\simeq7.4$. We stress two points of honesty.
First, the cosmological constant sits at $C=0$ while being severely disfavored
by naturalness arguments: coherence and naturalness here point in opposite
directions, which is exactly the distinction drawn in
Section~\ref{subsec:relation_other_priors} and a reminder that the two must not
be conflated. Second, the recent observational hints of an evolving dark-energy
equation of state \citep{DESI:2024mwx} do \emph{not} incur a coherence cost per
se: a $(w_0,w_a)$ history realized by a minimally coupled scalar remains at
$C=0$. Coherence penalizes the structural \emph{route} by which dynamics is
implemented, not the existence of dynamics. This structure reproduces the
qualitative ranking already standard in the modified-gravity literature
\citep{Clifton:2011jh,JoyceLombriserSchmidt:2016,Heisenberg:2018vsk}.

\subsubsection{Worked example: the $(w_0,w_a)$ plane and the DESI preference}
\label{subsec:w0wa}

\begin{figure}[t]
\centering
\includegraphics[width=0.9\textwidth]{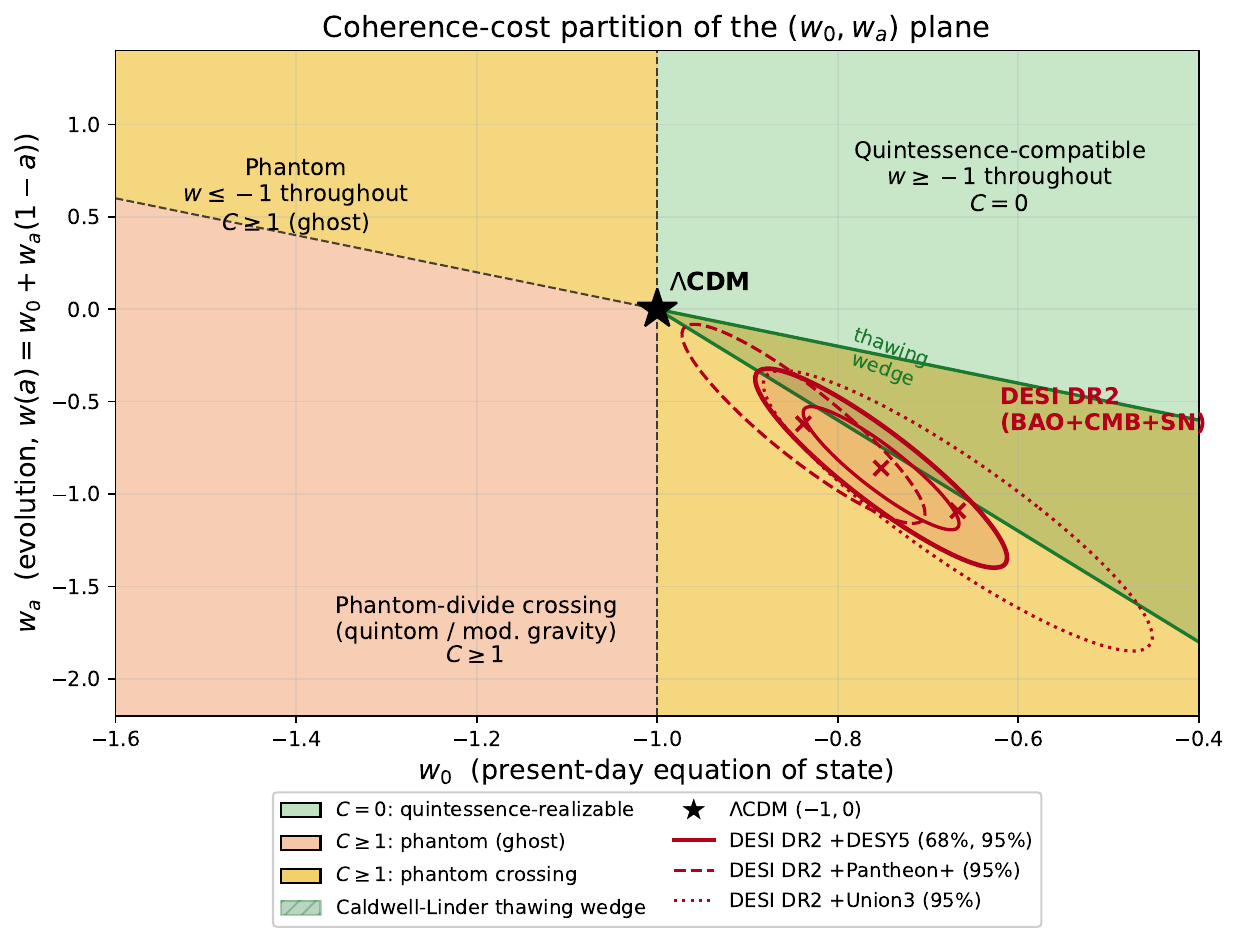}
\caption{Coherence-cost partition of the $(w_0,w_a)$ plane. \emph{Green:}
quintessence-realizable histories ($w\ge-1$ throughout, $C=0$), containing
$\Lambda$CDM at $(-1,0)$ and the Caldwell--Linder thawing wedge (hatched,
Eq.~\ref{eq:thawing}). \emph{Orange:} fully phantom histories ($w\le-1$
throughout), requiring a ghost ($C\ge1$). \emph{Yellow:} phantom-divide crossing
(Eq.~\ref{eq:crossing}), forbidden to a single minimally coupled scalar by
Vikman's no-go and hence requiring an extra degree of freedom or modified gravity
($C\ge1$). The actual DESI DR2 $w_0w_a$CDM confidence contours (BAO$+$CMB$+$SN;
68\% and 95\%; Gaussian reconstruction of the published marginalized constraints
\citep{DESI:2025dr2}) lie in the crossing zone for every supernova sample.}
\label{fig:w0wa}
\end{figure}

The Chevallier--Polarski--Linder (CPL) form $w(a)=w_0+w_a(1-a)$
\citep{ChevallierPolarski2001,Linder2003} is a fitting function, not a model, so
its coherence cost is that of the cheapest structure able to realize the history
while respecting the grammar $G(T_{\rm GR})$ of Eq.~(\ref{eq:GGR}); the resulting
cost is a lower bound on the structural price of a region. Since $w$ is monotonic
in $a$, its extrema are $w_0$ (today) and $w_0+w_a$ (early times), and at $a=1$
the Caldwell--Linder variable is $w'\equiv dw/d\ln a=-w_a$. Three structurally
distinct regions follow (Fig.~\ref{fig:w0wa}).

{\em Zero cost.} A canonical, minimally coupled scalar
\citep{RatraPeebles1988,Wetterich1988} obeys $w\ge-1$, so any history with
$w\ge-1$ throughout ($w_0\ge-1$ and $w_a\ge-(1+w_0)$) is realized at $C=0$,
respecting every entry of $G(T_{\rm GR})$. This region contains the cosmological
constant at $(w_0,w_a)=(-1,0)$ and the Caldwell--Linder thawing wedge
\citep{CaldwellLinder2005}, which via $w'=-w_a$ reads
\begin{equation}
-3(1+w_0)\;\le\;w_a\;\le\;-(1+w_0)\qquad(w_0\ge-1).
\label{eq:thawing}
\end{equation}
Crucially, \emph{dynamical} dark energy ($w_a\neq0$) is fully compatible with
$C=0$ as long as it stays on the non-phantom side of the divide.

{\em Nonzero cost.} A wholly phantom history ($w\le-1$ throughout) requires
a wrong-sign kinetic term \citep{Caldwell2002}, i.e.\ a ghost whose vacuum is
unstable to runaway pair production \citep{CarrollHoffmanTrodden2003,
ClineJeonMoore2004}; this violates the no-ghost/stability content of
$G(T_{\rm GR})$, so $C\ge1$. A history that \emph{crosses} the phantom divide,
\begin{equation}
(w_0+1)\,(w_0+w_a+1)<0,
\label{eq:crossing}
\end{equation}
cannot be produced by a single minimally coupled scalar without a perturbation
instability at the crossing (Vikman's no-go \citep{Vikman2005}); a stable crossing
needs an extra degree of freedom (the two-field quintom
\citep{FengWangZhang2005}), a non-minimal coupling, or modified gravity, each
adding a structural element beyond the $C=0$ scalar. Hence $C\ge1$, rising to
$C\simeq2$ when the realization both adds a field and breaks the equivalence
principle.

{\em The DESI preference.} The DESI baryon-acoustic data, with the CMB and
supernovae, prefer CPL over $\Lambda$CDM at $2.8$--$4.2\sigma$ depending on the
supernova sample; the DR2 marginalised constraints are
$(w_0,w_a)\simeq(-0.838,-0.62)$, $(-0.752,-0.86)$ and $(-0.667,-1.09)$ for the
Pantheon$+$, DES-Y5 and Union3 combinations \citep{DESI:2024mwx,DESI:2025dr2}.
Every one has $w_0>-1$ and $w_0+w_a<-1$, hence satisfies Eq.~(\ref{eq:crossing}):
the DESI confidence contours lie in the crossing zone (Fig.~\ref{fig:w0wa}), with
the divide crossed near $z\simeq0.4$. The coherence prior therefore does \emph{not} penalize evolving
dark energy---a thawing history obeying Eq.~(\ref{eq:thawing}) is as coherent as
$\Lambda$CDM---but it does penalize the phantom crossing that the best fit
implies. Writing $M_\times$ for a crossing realization ($C\simeq1$) and
$M_\Lambda$ for $\Lambda$CDM ($C=0$),
\begin{equation}
\frac{P(M_\times|D)}{P(M_\Lambda|D)}
=B_{\times\Lambda}\,e^{-\alpha(C_\times-C_\Lambda)}
\;\xrightarrow{\;\alpha=1\;}\;0.37\,B_{\times\Lambda},
\label{eq:w0wa_odds}
\end{equation}
so roughly one extra band of Jeffreys evidence is demanded of the data; against a
non-crossing thawing model ($C=0$) the prior is uniform and the comparison reverts
entirely to the likelihood. Since non-phantom fits remain compatible at
$\sim2\sigma$ and the CPL crossing may be a parametrization artifact of the
monotonic ansatz \citep{Zhao2017}, the coherence-informed reading is to treat the
DESI result as evidence for late-time dark-energy \emph{dynamics} rather than for
a genuine phantom crossing---a verdict the framework will revise if the crossing
becomes decisive. (We read the no-ghost/null-energy condition as the Wilsonian
content of the second-order principle in $G(T_{\rm GR})$, not as a separately
listed rule.)

\subsubsection{Inflationary model selection.} The background theory is single-field
slow-roll inflation \citep{Mukhanov2005}, whose grammar comprises single-clock
dynamics (a single adiabatic degree of freedom during inflation), near scale
invariance, near Gaussianity, adiabaticity of the primordial perturbations, the
Bunch--Davies initial vacuum, and a subluminal sound speed (locality and
causality). It is important that ``near scale invariance'' is part of the
grammar in its approximate form: the small, slow-roll-predicted departure from
exact scale invariance ($n_s\simeq0.965$) is a structural \emph{feature}, not a
violation, so a red tilt incurs no cost (it is also experimentally verified \cite{Spergel:2003cb,Planck:2018jri}). Single-field slow-roll models, for
example Starobinsky $R^2$ and plateau potentials, lie at $C=0$, as do
multi-field constructions that reduce to an effectively single adiabatic clock
once heavy fields are integrated out and the adiabatic limit is reached. A
genuinely multi-field model that leaves a detectable isocurvature component pays
$\delta=1$ on adiabaticity, $C=1$; reassuringly, the same models are also
disfavored by the tight Planck isocurvature bounds, so the coherence prior and
the Bayes factor act in concert rather than in tension. Non-Bunch--Davies
initial states likewise carry $C=1$. A strongly non-Gaussian single-field model
(for instance DBI inflation with a small sound speed) pays $\delta=1$ on near
Gaussianity \emph{unless} the non-Gaussian structure is required by an
independent ultraviolet or string completion, in which case the
independent-motivation clause sets $\delta=0$, a clean worked example of that
clause in action. Ghost or Lorentz-violating inflation breaks locality and
carries $C\ge1$. The Coherence Principle thereby organizes a model space that
has been informally ranked along these lines for two decades
\citep{MartinRingeval:2014,Planck:2018jri,Achucarro:2022qrl}.

\subsubsection{Beyond-Standard-Model extensions.} The background grammar is
$G(T_{\rm SM})$ of Eq.~(\ref{eq:GSM}). Type-I, II and III seesaw mechanisms
extend the gauge sector in a universality-preserving manner and break lepton
number through a controlled Majorana mass, so $C=0$, exactly as for the unified
model of Section~\ref{subsec:illustration_neutrino}. Supersymmetric extensions
such as the MSSM, and grand-unified theories such as $SU(5)$ and $SO(10)$,
preserve the gauge, Lorentz and renormalizability structure and extend the
particle content universally, so $C=0$; they predict testable signatures
(superpartners, proton decay). It is worth emphasizing that the
\emph{naturalness} arguments that originally motivated low-scale supersymmetry
are logically separate from coherence: the non-observation of superpartners at
the LHC constrains these models through the \emph{likelihood}, not through the
coherence prior, and conflating the two would misattribute an empirical null
result to a structural one. Models that introduce a family-dependent $Z'$ or
otherwise non-universal gauge couplings, proposed at various times to account
for flavour anomalies, violate family universality $p_2$; thus carry $C=1$ unless
a robust anomaly independently demands the violation. The lepton-universality
hints in $b\to s\ell\ell$ transitions that motivated several such models have
since regressed substantially toward the Standard Model
\citep{ParticleDataGroup:2024,LHCb:2022qnv,Buttazzo:2017ixm}; we present this only
as an illustration, not a proof, that the coherence prior's standing penalty on
unmotivated universality violation was well placed. Explicitly non-renormalizable
extensions lacking a clear ultraviolet completion incur $\delta=1$ on $p_5$.

\subsubsection{Statistical and astrophysical inference.} The principle applies
equally to inferences whose background theory is not fundamental but
phenomenological. A representative case is the perturbative effective field
theory of galaxy bias used to interpret large-scale-structure surveys, for which
$T_{astro}$ is the set of theories that  describe the observable Universe at astrophysical/cosmological scales.  This includes, GR, atomic and nuclear physics and hydrodynamics. The grammar  $G(T)$ comprises among other things, the locality of the
bias expansion (galaxy statistics depend on the local long-wavelength fields and
their spatial derivatives), the equivalence-principle constraints encoded in the
large-scale consistency relations (no dependence on the absolute gravitational
potential, and the absence of significant large-scale velocity bias at late times), the derivative and
renormalizability ordering of the bias expansion (a finite operator basis at
each order), statistical isotropy and homogeneity, and Poisson shot noise as the
leading stochasticity \citep{Desjacques:2016bnm,Heavens:2018adv}. Candidate
``models'' here are competing nuisance-parameter schemes. A scheme that
introduces an ad hoc scale-dependent bias term lying outside the
symmetry-allowed derivative expansion incurs $\delta=1$, so the coherence cost
penalizes nuisance freedoms that are neither calibrated against simulations nor
permitted by the effective-theory grammar (see e.g., \cite{NovellMasot:2023dfg,DESI:2024mwx}). The logic is identical to the
fundamental-physics cases; only the status of the background theory differs.

\begin{table}[t]
\centering
\renewcommand{\arraystretch}{1.25}
\begin{tabular}{>{\raggedright\arraybackslash}p{0.30\textwidth}
                >{\raggedright\arraybackslash}p{0.42\textwidth}
                >{\centering\arraybackslash}p{0.07\textwidth}
                >{\centering\arraybackslash}p{0.10\textwidth}}
\hline
\textbf{Candidate model} & \textbf{Grammar principle at stake} & $C$ & $e^{-\alpha C}$\\
\hline
\multicolumn{4}{l}{\emph{Dark energy / modified gravity}, $T=T_{\rm GR}$}\\
$\Lambda$, minimal quintessence & none violated & $0$ & $1$\\
$k$-essence (superluminal) & local causality / analyticity & $1$ & $0.37$\\
Screened scalar--tensor, $f(R)$, Horndeski & equivalence principle ($p_2^{g}$) & $1$ & $0.37$\\
Massive / bimetric gravity & tensor structure, ghost ($p_3^{g},p_4^{g}$) & $1$--$2$ & $0.37$--$0.14$\\
Lorentz-violating / nonlocal gravity & locality, Lorentz ($p_4^{g}$) & $2$--$3$ & $0.14$--$0.05$\\
\hline
\multicolumn{4}{l}{\emph{Inflation}, $T=$ single-field slow roll}\\
Single-field slow roll; adiabatic multi-field & none violated & $0$ & $1$\\
Surviving isocurvature; non-Bunch--Davies & adiabaticity / vacuum & $1$ & $0.37$\\
Strong non-Gaussianity (unmotivated) & near Gaussianity & $1$ & $0.37$\\
\hline
\multicolumn{4}{l}{\emph{Beyond Standard Model}, $T=T_{\rm SM}$}\\
Seesaw, MSSM, $SU(5)$/$SO(10)$ GUT & none violated & $0$ & $1$\\
Non-universal $Z'$ (unmotivated) & family universality ($p_2$) & $1$ & $0.37$\\
Non-renormalizable EFT (no UV completion) & renormalizability ($p_5$) & $1$ & $0.37$\\
\hline
\end{tabular}
\caption{Illustrative coherence costs and prior weights ($\alpha=1$) for the
candidate models discussed in Section~\ref{subsec:general_apps}. Costs depend on
the stated grammar and on the independent-motivation clause; ranges indicate the
spread across sub-cases. The table is intended to make the assignments explicit
and contestable, not to fix them definitively.}
\label{tab:coherence_costs}
\end{table}

In each of these settings, the workflow of Section~\ref{subsec:workflow}
remains unchanged. The only application-dependent inputs are the grammar
$G(T)$ and the catalogue of candidate models.

\subsection{Historical case studies: retrodicting paradigm shifts}
\label{subsec:historical}

A prior prescription that claims to formalize the structural reasoning of
physicists should be tested against the historical record, where the outcomes
are already known. 
 When applied to past historical cases, would the Coherence Principle have favored the choice that history later
proved successful?
We examine four episodes. The exercise is unavoidably
somewhat postdictive and therefore exposed to hindsight bias. We mitigate this by imposing a time-ordering requirement:  the grammar
$G(T)$ is fixed using only structural rules that were empirically validated
before the episode under discussion, and not rules inferred from its eventual
outcome.

Then the rules of Section~\ref{subsec:workflow} are applied 
mechanically. The four cases are chosen because each stresses a different
feature of the framework: the time-dependence of the grammar (Einstein versus
Newton), a clean victory for coherence in the weak-data regime (Pauli's
neutrino), and an instructive apparent failure (parity violation).

\subsubsection{Would the Coherence Principle have favored Einstein over
Newton?}
\label{subsubsec:einstein}

This is the sharpest test, and the answer is instructive precisely because it
depends entirely on the date at which the grammar is fixed.

\emph{Pre-1905 anchoring.} If $T$ is Newtonian mechanics and universal
gravitation as validated up to about 1900, then
\begin{align}
G(T_{\rm N})=\{\,
& \text{absolute time and simultaneity},\;\text{Galilean invariance},\;
\text{Euclidean space},\nonumber\\
& \text{instantaneous action at a distance},\;\text{determinism}\,\}.
\end{align}
A nascent general-relativistic theory violates the first four entries: it
abolishes absolute simultaneity, replaces Galilean by Lorentz invariance, makes
spatial geometry dynamical and curved, and propagates gravitational influence at
the finite speed $c$. Applied mechanically, $C(\text{GR}\,|\,T_{\rm N})\simeq4$,
a prior suppression $e^{-4}\simeq0.018$. A coherence prior anchored to 1900
physics would therefore have \emph{strongly disfavored} general relativity. We
state this plainly: applied to a grammar that was about to be superseded, the
principle is conservative and would have backed Newton.

\emph{Post-1905 anchoring.} By 1915, however, the validated grammar had already
moved. Special relativity \citep{Einstein:1905} was, within a decade, an
empirically established background theory in its own right, supported by the
Michelson--Morley null result, the electrodynamics of moving bodies, and the
relativistic energy--momentum relation tested in early particle experiments (see Sec.\ref{subsubsec:lorentz} for Coherence principle application in this context). Therefore 
the grammar of 1915 included
\begin{align}
G(T_{\rm SR})=\{\,
& \text{Lorentz invariance},\;\text{finite maximal signal speed }c,\nonumber\\
& \text{no absolute simultaneity},\;\text{local causality}\,\}.
\end{align}
Against \emph{this} grammar it is the incumbent that is in difficulty.
Newtonian gravity, through its instantaneous action at a distance, violates
Lorentz invariance and the finite-signal-speed and causality principles
outright, so $C(\text{Newtonian gravity}\,|\,T_{\rm SR})\ge2$. General
relativity, by contrast, was constructed precisely to be locally Lorentz
invariant and causal, and to embed the equivalence principle---the empirical
equality of inertial and gravitational mass, verified to high precision by
E\"otv\"os \citep{Eotvos:1922}---as a structural feature rather than an
unexplained coincidence; hence $C(\text{GR}\,|\,T_{\rm SR})\simeq0$. With
$\alpha=1$ the prior odds now favor Einstein,
\begin{equation}
\frac{P(\text{GR}\,|\,T_{\rm SR})}{P(\text{Newton}\,|\,T_{\rm SR})}
= e^{\,C_{\rm Newton}-C_{\rm GR}}\simeq e^{2}\simeq7,
\end{equation}
\emph{before} the decisive empirical tests are brought to bear. Those tests then
point the same way: the anomalous advance of Mercury's perihelion of
$43''$ per century, unresolved within Newtonian gravity since Le~Verrier in 1859,
was accounted for exactly by general relativity in 1915
\citep{Einstein:1915perihelion,Einstein:1915fieldeq}, and the $1.75''$
deflection of starlight---twice any Newtonian corpuscular estimate---was
confirmed at the 1919 solar eclipse \citep{DysonEddingtonDavidson:1920}. The
Bayes factor and the coherence prior reinforce one another and the posterior is
overwhelming.

Three lessons follow. \emph{(i) Time-stamping is mandatory.} The verdict is only
as reliable as the grammar it is anchored to, and the correct anchor is the most
secure \emph{currently} validated grammar---here special relativity, not
Newtonian mechanics. Anchoring to a superseded grammar is the dominant way to
misuse the principle, and it is exactly the time-dependence caveat of
Section~\ref{subsec:limitations}. \emph{(ii) The principle is not status-quo
conservatism.} Once special relativity was validated, the principle turned
\emph{against} the older incumbent; a framework that merely entrenched the
dominant theory of the day could not do this. \emph{(iii) Explanatory credit
suggests an extension.} General relativity does more than avoid violating the
equivalence principle: it \emph{explains} a previously unexplained empirical
coincidence (the equality of inertial and gravitational mass). The binary
indicator $\delta_i\in\{0,1\}$ of Eq.~(\ref{eq:delta_indicator}) cannot represent
this. It motivates a \emph{signed} generalization of the coherence cost, in which
a principle may be violated, respected, or positively \emph{explained} by a
candidate model, the last earning a negative cost: a ``coherence bonus''. We
develop this extension formally, with guardrails against abuse and a worked
treatment of the present case, in Section~\ref{subsec:signed}; the
Einstein--Newton comparison is its cleanest motivation.

Finally, we are careful not to overclaim. This is a reconstruction: Einstein was
not guided by a Bayesian prior, and the historical acceptance of general
relativity was driven by its empirical successes. The claim is only that, \emph{had}
a coherence prior been computed in 1915 against the then-validated grammar, it
would have favored general relativity, and that computing it against the wrong,
pre-1905 grammar would have given the opposite answer. The episode is thus as
much a warning about anchoring as it is a success story.

\subsubsection{Pauli's neutrino versus abandoning energy conservation}
\label{subsubsec:pauli}

In 1930 the continuous energy spectrum of nuclear $\beta$ decay confronted
physics with a stark choice. One response, developed by Bohr from the earlier
Bohr--Kramers--Slater proposal of merely statistical conservation in radiative
processes \citep{BohrKramersSlater:1924}, was to abandon microscopic energy
conservation in $\beta$ decay and retain it only on the average; call this model
$M_{\rm Bohr}$. The other, proposed by Pauli in his celebrated letter of 4
December 1930 to the Tübingen conference \citep{Pauli:1930,Brown:1978}, was to
postulate a new neutral, light, spin-$\tfrac12$, weakly interacting particle
(later named the neutrino by Fermi \citep{Fermi:1934}) that carries away the
missing energy, momentum and angular momentum, preserving every conservation law
exactly; call this $M_{\rm Pauli}$. The relevant grammar of 1930 physics is
\begin{align}
G(T)=\{\,
& \text{energy conservation},\;\text{momentum conservation},\;
\text{angular-momentum conservation},\nonumber\\
& \text{spin--statistics},\;\text{Lorentz invariance}\,\}.
\end{align}
Energy conservation had, in Pauli's own words, ``stood the test in all fields of
physics.'' The model $M_{\rm Bohr}$ violates energy, momentum and
angular-momentum conservation simultaneously, and stood in tension with the
spin--statistics of nuclei such as $^{14}$N (the ``nitrogen anomaly'' that Pauli
also sought to resolve): $C(M_{\rm Bohr})\simeq3$. The model $M_{\rm Pauli}$
respects every principle, since positing a new particle \emph{within} the
existing symmetry and conservation structure is not a grammar violation---in
direct analogy with the unified neutrino-mass model of
Section~\ref{subsec:illustration_neutrino}---so $C(M_{\rm Pauli})=0$. With
$\alpha=1$ the prior odds are $P(M_{\rm Pauli})/P(M_{\rm Bohr})=e^{3}\simeq20$,
or $e^{1}\simeq2.7$ if only energy conservation is counted; either way coherence
favors Pauli.

This is the ideal regime for the framework. In 1930 the data could not
discriminate between the two models---the neutrino was not detected directly
until Reines and Cowan in 1956 \citep{ReinesCowan:1956}---so the Bayes factor was
near unity and the \emph{prior} carried the inference. History vindicated Pauli
decisively. The case also makes concrete the separation, argued in
Section~\ref{subsec:relation_other_priors}, between coherence and Occam
parsimony: counting entities, Occam mildly \emph{disfavors} Pauli, who
introduced a new and then-undetectable particle, whereas the coherence prior
strongly \emph{favors} him for preserving the conservation grammar. The two
criteria pull in opposite directions, coherence prevails, and conflating the two
would have given precisely the wrong steer.

\subsubsection{Parity violation: an instructive apparent failure}
\label{subsubsec:parity}

By the mid-1950s the conservation of parity ($\mathcal{P}$) was widely treated as a
structural rule, having held without exception in strong and electromagnetic
processes. The $\theta$--$\tau$ puzzle---two particles of equal mass and lifetime
decaying into final states of opposite parity---admitted two readings: either
$\theta$ and $\tau$ are distinct particles and $\mathcal{P}$ is conserved, or they are a
single particle (the $K$ meson) and $\mathcal{P}$ is violated. A naive application of the
Coherence Principle, using a grammar that \emph{included} parity conservation as
a universal principle, would assign a coherence cost to any parity-violating
proposal and would therefore have favored ``$\theta\neq\tau$'' and disfavored
the hypothesis of Lee and Yang \citep{LeeYang:1956}. The 1957 $^{60}$Co
experiment of Wu and collaborators \citep{Wu:1957} then revealed maximal parity
violation in the weak interaction: the naive coherence verdict was simply wrong.

The resolution is the real lesson. Lee and Yang's actual 1956 argument was that
parity conservation had \emph{never been experimentally tested} in the weak
interactions; its assumption there was an unjustified extrapolation from the
strong and electromagnetic sectors. In the language of the Coherence Principle,
the principle ``parity conservation'' was empirically validated only within the
strong/electromagnetic grammar, and its inclusion in the grammar of the weak
sector was an untested assumption. A correctly constructed, \emph{domain-resolved}
grammar assigns $\delta=0$ to a parity-violating weak-interaction model, because
the principle is not a validated rule of that sector; equivalently, the
$\theta$--$\tau$ puzzle itself furnished exactly the independent motivation that
the indicator of Eq.~(\ref{eq:delta_indicator}) requires for setting $\delta\to0$.
The failure is therefore not of the principle but of an over-extended grammar.

Three lessons follow. \emph{(i) The grammar must be domain-resolved.} A rule
validated in one sector is not automatically a rule of another, and the
over-inclusion of unvalidated principles is the dominant failure mode of any
coherence-style reasoning. \emph{(ii)} This is precisely why the framework
demands that every $p_i$ be a repeatedly tested rule of the relevant domain and
why it carries the revisability and independent-motivation clauses of
Sections~\ref{subsec:properties} and~\ref{subsec:limitations}. \emph{(iii)
Calibration consistency.} Parity is one of the historically resolved cases in
which the grammar-violating model prevailed, in the calibration band
$f\sim0.2$--$0.4$ of Eq.~(\ref{eq:alpha_calibration}); a dogmatic prior
($\alpha\to\infty$) would have rendered the inference unfalsifiable and wrong,
whereas a moderate $\alpha\sim1$ is straightforwardly overturned by Wu's decisive
data, exactly as the falsifiability property of
Section~\ref{subsec:properties} requires.

\subsubsection{Special relativity versus the Lorentz aether: the prior decides}
\label{subsubsec:lorentz}

The three episodes above all involve a Bayes factor that is either decisive
(Mercury, the eclipse, the $^{60}$Co experiment) or merely deferred (the
neutrino). The displacement of the Lorentz aether theory by special relativity
is different and, for our purposes, especially clean: around 1905 the two
frameworks were \emph{empirically degenerate}. Lorentz's electron theory, with
its contraction hypothesis and local time \citep{Lorentz:1904}, reproduced the
same observable predictions as Einstein's special relativity
\citep{Einstein:1905}, including the null result of the Michelson--Morley
experiment \citep{MichelsonMorley:1887}. With $B\simeq1$, the likelihood is
silent and the inference is carried entirely by the prior---precisely the regime
the Coherence Principle is designed for.

The relevant validated grammar by 1905 already contained the principle of
relativity (no preferred inertial frame), a rule repeatedly confirmed in
mechanics since Galileo and reinforced for electrodynamics by the persistent
failure to detect any motion relative to the aether. Lorentz's theory retains a
preferred rest frame---the aether---rendered undetectable \emph{by construction}
through the contraction hypothesis; it thereby violates the principle of
relativity in the structural sense while evading it observationally, an
unmotivated complication that earns $\delta=1$ on that principle, so
$C(\text{Lorentz})=1$. Special relativity instead elevates the principle of
relativity to a postulate and \emph{derives} length contraction, time dilation
and the Lorentz covariance of all physics as consequences, so
$C(\text{SR})=0$. With $\alpha=1$ the prior odds favor special relativity by
$e^{1}\simeq2.7$.
With the coherence bonus
for converting the previously coincidental Lorentz invariance of Maxwell's
equations into a structural theorem (see the signed extension of
Section~\ref{subsec:signed}), the preference strengthens to
$e^{2}\simeq7$. History abandoned the aether.

For transparency, we highlight two points where the proposal is deliberately conservative.
First, the equivalence was always observational, not
logical, and some authors have regarded the choice as conventional; the
Coherence Principle does not pretend to settle that philosophical question, only
to make explicit which structural consideration tips a degenerate inference and
why. Second, the later embedding of Lorentz invariance in general relativity and
quantum field theory provided strong subsequent validation of the
no-preferred-frame grammar, but that is hindsight: the point here is that the
\emph{1905} coherence prior already favored special relativity before those
developments, on the basis of the grammar then in force.

\paragraph{Synthesis}

Across the four episodes, the Coherence Principle favors the historically
successful option when it is applied with a domain-resolved, correctly
time-stamped grammar and the independent-motivation clause. This is the case for
Pauli's neutrino, Einstein under the post-1905 anchor, and special relativity
over the aether in the degenerate-data regime. The apparent failure in the
parity case is 
instead a failure of input specification: the assumed grammar had
been extended beyond its empirical support.

The examples also expose the main misuse modes of the framework. One may anchor
the grammar to a superseded theory, as in the pre-1905 Einstein--Newton reading,
or include principles that were not yet empirically validated, as in the naive
parity reading. Because these choices are explicit inputs, the failure modes are
diagnosable rather than hidden.

The aether case adds a final lesson. When the data are degenerate
($B\simeq 1$), the coherence prior is not merely a tiebreaker of last resort. It
can be the decisive contribution, and the coherence bonus (the signed extension below) makes this even
sharper.

\subsection{A signed extension: explanatory credit and the coherence bonus}
\label{subsec:signed}

The historical case studies, and the Einstein--Newton comparison in particular,
expose a limitation of the binary indicator of Eq.~(\ref{eq:delta_indicator}): it
records whether a model \emph{violates} a structural, empirically validated, principle, but not whether a
model positively \emph{explains} one. General relativity does not merely refrain
from violating the equivalence principle; it derives the equality of inertial and
gravitational mass, an unexplained empirical coincidence in Newtonian
gravity, as a structural theorem. Special relativity likewise converts the
Lorentz invariance of Maxwell's equations from an accident into a consequence of
its postulates. A framework that scores both theories at $C=0$ misses a genuine
structural distinction. We therefore introduce a conservative \emph{signed}
generalization of the coherence cost.

\paragraph{Definition.} Replace the indicator $\delta_i\in\{0,1\}$ by a signed
\emph{structural valuation}
\begin{equation}
\sigma_i(M,T)=
\begin{cases}
+1, & M \text{ violates } p_i \text{ without independent motivation},\\
\phantom{+}0, & M \text{ respects } p_i \text{ as an assumption},\\
-1, & M \text{ explains } p_i \text{ (derives it from deeper posited structure)},
\end{cases}
\label{eq:sigma_indicator}
\end{equation}
and define the signed coherence cost and prior
\begin{equation}
C_{\pm}(M|T)=\sum_{i=1}^{n} w_i\,\sigma_i(M,T)\in\mathbb{Z},
\qquad
P(M|T)\propto \exp\!\left[-\alpha\,C_{\pm}(M|T)\right].
\label{eq:signed_cost}
\end{equation}
The maximum-entropy and multiplicative-factorization arguments of
Section~\ref{subsec:coherence_cost} are unaffected, since they constrain only the
functional form of the map and not the sign of its argument; the sole change is
that the domain of the cost is enlarged from the non-negative integers to all of
$\mathbb{Z}$. When no explanatory credits are assigned ($\sigma_i\in\{0,1\}$),
$C_{\pm}=C$ and the original construction is recovered exactly, so the extension
is strictly a refinement.

\paragraph{Guardrails.} Explanatory credit is more contestable than the
recording of a violation, and an undisciplined ``explanation'' indicator would
reintroduce precisely the aestheticism the framework is meant to exclude. We
therefore admit $\sigma_i=-1$ only when all of the following hold.
(G1)~\emph{derivation}: $p_i$ follows as a theorem in $M$ from assumptions that
are strictly more fundamental than, or logically independent of, $p_i$ itself;
(G2)~\emph{no compensating violation}: $M$ does not purchase the explanation by
violating another principle, so that an explanatory credit cannot offset an
unrelated structural pathology elsewhere in $C_{\pm}$;
(G3)~\emph{consensus on the feature to be explained  
}: there is inter-subjective agreement
that $p_i$ was a previously unexplained regularity in $T$ rather than an
already-derived consequence; and (G4)~\emph{independent testability}: the deeper
structure invoked carries empirical consequences beyond $p_i$ itself, so the
credit is predictive rather than purely retrospective. These conditions are
deliberately stringent: the burden of proof for a bonus is higher than for a
penalty.

\paragraph{Worked examples.} In Newtonian gravity the weak equivalence principle
is an unexplained input (the empirical equality $m_{\rm i}=m_{\rm g}$, verified by
E\"otv\"os \citep{Eotvos:1922}); in general relativity it is a consequence of the
identification of free fall with geodesic motion, which is independent of the
falling body's mass. All four guardrails are met---G4 in particular, since the
geometric structure also predicts light bending and the perihelion advance---so
$\sigma_{\rm EP}(\text{GR})=-1$ and, against the post-1905 grammar of
Section~\ref{subsubsec:einstein}, $C_{\pm}(\text{GR}\,|\,T_{\rm SR})=-1$ versus
$C_{\pm}(\text{Newton}\,|\,T_{\rm SR})\ge+2$, sharpening the prior odds from
$e^{2}$ to $e^{3}\simeq20$. Similarly, special relativity earns
$\sigma=-1$ for deriving universal Lorentz covariance
(Section~\ref{subsubsec:lorentz}), strengthening its advantage over the aether
theory from $e^{1}$ to $e^{2}$. We recommend, as a matter of practice, that both
the unsigned cost $C$ and the signed cost $C_{\pm}$ be reported, with the
difference $C-C_{\pm}$ treated as an explicit sensitivity band that isolates the
contribution of the more contestable explanatory credits.

\subsection{Limitations and methodological caveats}
\label{subsec:limitations}

The Coherence Principle is not a 
 general-purpose solution 
and should be applied with
the caveats appropriate to any prior prescription.

\emph{i)} The grammar $G(T)$ is itself empirical and may evolve. A
principle currently regarded as a structural rule may be invalidated by future
data; when this happens, the corresponding entry must be removed from $G(T)$
or reassigned $\delta_i=0$ for the relevant model. The Coherence Principle is
therefore explicitly time-dependent at the level of $G(T)$. We see this   as a 
methodological feature rather than a bug.

\emph{ii)} The coherence prior is a model-space prior and does not replace
the parameter-space prior $\pi(\theta|M)$ that enters within the evidence
integral. Both must be specified, and the standard tools of objective Bayesian
analysis (Jeffreys priors, reference priors, hierarchical priors) remain
appropriate for $\pi(\theta|M)$~\citep{Bernardo:1979,Heavens:2018adv}.

\emph{iii)} The framework cannot adjudicate between models that share the
same $C$. In such cases, the coherence prior is uniform, and discrimination
must come from the Bayes factor alone. This is a feature: the Coherence
Principle is designed to formalize structural preferences and remains silent
when no structural distinction is available.

\emph{iv)} In regimes where the data are decisive
($B_{12}\gg \Pi_{12}^{-1}$), the inference is dominated by the likelihood and
the coherence prior has only a negligible effect on the posterior. In such
regimes, the coherence prior plays a verification role rather than a steering
role: it provides a consistency check, not a tiebreaker.

\emph{v)} The framework deliberately avoids quantifying purely aesthetic
criteria (elegance, mathematical beauty) that are sometimes invoked in
theoretical physics. Such criteria are not part of $G(T)$ unless they can be
restated as empirically--validated structural rules.

\emph{vi)--grammar 
discipline.} The historical case studies of
Section~\ref{subsec:historical} expose two failure modes that are not failures of
the construction but of its inputs, and both are worth stating as explicit
prescriptions because both are diagnosable before any cost is computed. The first
is \emph{anchoring to a superseded grammar}: a coherence prior is only as
trustworthy as the theory it is anchored to, and the correct anchor is always the
most secure \emph{currently} validated grammar. Computed against pre-1905
Newtonian mechanics, the principle disfavors general relativity; computed against
the special-relativistic grammar already in force by 1915, it favors it. The
remedy is to \emph{time-stamp} $G(T)$ and to re-anchor it whenever a deeper
background theory is validated. The second failure mode is
\emph{over-extension of the grammar}: importing a principle validated in one
domain into another where it has never been tested, as parity conservation was
silently carried from the strong and electromagnetic sectors into the weak
interaction. The remedy is to \emph{domain-resolve} $G(T)$, admitting each $p_i$
only for the sectors in which it has been repeatedly tested, and to treat an
untested extrapolation as absent from the grammar (so $\delta_i=0$) rather than
as a validated rule (alternatively this can be tuned by suitable choice of the weights). Time-stamping and domain-resolution together constitute the
minimal hygiene required of any grammar before it is used, and we recommend that
both be documented explicitly alongside the list $G(T)$ in any application.

\subsection{Discussion and outlook}
\label{subsec:coherence_discussion}

A long-standing question, especially acute in cosmology and theoretical physics, lies at the
interface between Bayesian statistics and theory-driven science:
how can
the structural lessons of successful theories be incorporated into
quantitative Bayesian inference without lapsing into aestheticism?

The proposed Coherence Principle answers by encoding those lessons as a coherence cost on candidate models, mapped to a
prior through the exponential of Eq.~(\ref{eq:coherence_prior}).

The
construction is transparent because every input is listed; reproducible
because every input can be challenged and varied; and falsifiable because
the prior odds it induces are finite and overruled by sufficiently strong
data.

\begin{table}[tbp]
\centering
\small
\setlength{\tabcolsep}{4pt}
\renewcommand{\arraystretch}{1.05}
\begin{tabular}{|l|l|l|}
\hline
Discipline & Background theory or framework & Example models \\
\hline
\hline
Biology & Evolutionary theory & Population-genetic models \\
\hline
Economics & Rational-choice or & Market, inflation, \\
 & Behavioral frameworks & or decision models \\
\hline
Epidemiology & Transmission theory & SIR, SEIR, or agent-based models \\
\hline
Climate science & Physical climate theory & Climate models and parameterizations \\
\hline
Neuroscience & Theories of perception & Computational models \\
 & or cognition & of neural processing \\
\hline
Ecology & Ecosystem theory and & Predator--prey, metapopulation, \\
 & population dynamics & or niche models \\
\hline
Earth science & Plate tectonics & Mantle-convection, subduction-zone, \\
 & & or earthquake-risk models \\
\hline
Psychology & Cognitive and & Decision, memory, attention, \\
 & Behavioral theory & or learning models \\
\hline
Finance & No-arbitrage and & Asset-pricing, risk, volatility, \\
 & market-efficiency principles & or portfolio models \\
\hline
Political science & Collective-choice and & Voting, coalition-formation, \\
 & institutional theory & or conflict models \\
\hline
Sociology & Social-network and & Diffusion, segregation, \\
 & social-structure theory & or opinion-dynamics models \\
\hline
Linguistics & Grammar and & Syntactic, semantic, \\
 & language-acquisition theory & or language-evolution models \\
\hline
Medicine & Disease-mechanism and & Diagnostic, prognostic, \\
 & pathophysiological theory & or treatment-response models \\
\hline
Public health & Causal and & Intervention, risk-factor, \\
 & population-health frameworks & or burden-of-disease models \\
\hline
Machine learning & Statistical learning theory & Bayesian, neural-network, kernel, \\
 & & or causal models \\
\hline
\end{tabular}
\caption{Examples of theory-informed model comparison beyond physics where the Coherence Principle may apply.}
\label{tab:theory-model-examples}
\end{table}

The neutrino hierarchy provides a particularly clean illustration of this
machinery because the only nontrivial structural rule at stake, family
universality, admits a binary indicator and because the relevant background
theory is the Standard Model in its most secure form. The same machinery
applies, with minor adaptation, to the comparison of dark-energy parametrizations,
inflationary models, modified-gravity extensions, beyond-Standard-Model
sectors and a wide class of cosmological inferences. In every case, the
Coherence Principle delivers an explicit \emph{epistemic-paradigm prior}
that quantifies the rational confidence assigned to structural rules
inferred from the accumulated success of a scientific paradigm.

The principle does not replace the Bayesian evidence and does not double-count
the Occam factor. It complements both. The evidence weights the data against
the prior parameter volume; the Occam factor it contains penalizes
parametric extravagance; and the Coherence Principle, operating one level
up, weights candidate \emph{models} by their structural compatibility with
empirically validated theoretical grammar. 

Future applications should explore
the calibration of $\alpha$ on extended historical ensembles, the systematic
construction of grammars beyond the Standard Model, and the integration of
the coherence prior with hierarchical and graphical-model representations of
the model space. The broader goal is to convert what has long been an
informal element of physical reasoning, confidence in the persistence of
validated structural rules, into a routine component of cosmological and
particle-physics inference, on the same methodological footing as the
likelihood and the parameter-space prior.

Although our examples are drawn from physics, the Coherence Principle is a general proposal for theory-informed inference  
across empirical sciences  and 
quantitative fields (Table~\ref{tab:theory-model-examples}).

\begin{acknowledgments}
This work was supported by a grant from the Simons Foundation (00017375, RJ).
Funding for the work of RJ and LV was partially provided by project PID2022-141125NB-I00,
and the “Center of Excellence Maria de Maeztu 2025-2029” award to the ICCUB funded by
grant CEX2024-001451-M from AEI/10.13039/501100011033.
\end{acknowledgments}

\bibliographystyle{JHEP}

\providecommand{\href}[2]{#2}\begingroup\raggedright\endgroup

\end{document}